\documentclass[conference, 10pt]{IEEEtran} 
\usepackage[normalem]{ulem}
\usepackage{dsfont}
\usepackage{amsmath,amssymb,amsfonts}
\usepackage{algorithmic}
\usepackage{graphicx}
\usepackage{textcomp}
\usepackage{caption}
\usepackage{subcaption}
\usepackage{latexsym}
\usepackage{amsmath}
\usepackage[mathscr]{euscript}
\usepackage[linesnumbered, algo2e, ruled,norelsize]{algorithm2e}
\SetArgSty{textnormal}
\usepackage{threeparttable}
\usepackage{threeparttablex}
\usepackage{commath}
\usepackage{mathtools}
\usepackage{slashbox}
\usepackage{pict2e}
\usepackage{epstopdf}
\usepackage{verbatim}
\usepackage{subfloat}
\usepackage{tabu}
\usepackage{booktabs}
\usepackage{float}
\usepackage{amssymb}
\usepackage{url}
\usepackage{multirow}
\usepackage{graphicx}
\usepackage{mathptmx}
\usepackage{cite}
\usepackage{footnote}
\usepackage{array}
\usepackage[table]{xcolor}
\usepackage{tabularx}
\usepackage{ulem}
\usepackage{algorithm}
\usepackage{amsthm} 
\allowdisplaybreaks
\def\BibTeX{{\rm B\kern-.05em{\sc i\kern-.025em b}\kern-.08em
    T\kern-.1667em\lower.7ex\hbox{E}\kern-.125emX}}
\usepackage[left=1.62cm,right=1.62cm,top=1.8cm]{geometry}
\allowdisplaybreaks

\begin{document}
\title{Joint Design of Probabilistic Constellation Shaping and Precoding for Multi-user VLC Systems}
\author{\IEEEauthorblockN{Thang K. Nguyen\IEEEauthorrefmark{1}, Thanh V. Pham\IEEEauthorrefmark{2}, Hoang D. Le\IEEEauthorrefmark{1}, Chuyen T. Nguyen\IEEEauthorrefmark{3}, and Anh T. Pham\IEEEauthorrefmark{1}}

\IEEEauthorblockA{\IEEEauthorrefmark{1}Computer Communications Lab., The University of Aizu, Japan.}

\IEEEauthorblockA{\IEEEauthorrefmark{2}Department of Mathematical and Systems Engineering, Shizuoka University, Japan.}

\IEEEauthorblockA{\IEEEauthorrefmark{3}School of Electrical and Electronic Engineering, Hanoi University of Science \& Technology, Vietnam}

}

\maketitle 
\begin{abstract}
This paper proposes a joint design of probabilistic constellation shaping (PCS) and precoding to enhance the sum-rate performance of multi-user visible light communications (VLC) broadcast channels subject to signal amplitude constraint. In the proposed design, the transmission probabilities of bipolar $M$-pulse amplitude modulation ($M$-PAM) symbols for each user and the transmit precoding matrix are jointly optimized to improve the sum-rate performance. The joint design problem is shown to be a complex non-convex problem due to the non-convexity of the objective function. To tackle the problem, the firefly algorithm (FA), a nature-inspired heuristic optimization approach, is employed to solve a local optima to the original non-convex optimization problem. The FA-based approach, however, suffers from high computational complexity. Therefore, we propose a low-complexity design based on zero-forcing (ZF) precoding, which is solved using an alternating optimization (AO) approach. Simulation results reveal that the proposed joint design with PCS significantly improves the sum-rate performance compared to the conventional design with uniform signaling. Some insights into the optimal symbol distributions of the two joint design approaches are also provided.
\end{abstract}

\vspace{-0.3 cm}
\section{Introduction}
The massive growth in Internet applications, especially multimedia ones that demand increasingly high data-rate transmission, has prompted rapid progress in research and development of wireless systems over the past decade. Due to its high data rate, license-free spectrum, and immunity to radio interference, visible light communications (VLC) is a promising technology to complement the coexisting radio frequency (RF) systems \cite{Matheus2019}. 

While RF communications allow for input signals with unbounded amplitudes under average power constraints, in VLC systems, an amplitude constraint is applied to the channel input to ensure the limited linear range due to LED peak power constraint and non-negativity signals \cite{PLS_2014}. According to \cite{Capacity_peak_1971} and \cite{GM_2016}, the capacity-achieving input distribution for a scalar Gaussian channel under an amplitude constraint is discrete with a finite number of symmetric mass points. While optimizing the input position, which is known as geometric constellation shaping, is generally impractical since it increases the transceiver complexity, shaping the input transmission probability, a.k.a probabilistic constellation shaping (PCS), is practically feasible with the use of a distribution matcher, such as the constant composition distribution matching (CCDM) \cite{CCDM_2016}.  

In fact, due to its low complexity and flexibility, PCS has been widely studied in optical fiber \cite{PCS_fiber_2019, PS_GS_2019} and optical wireless communications \cite{PS_VLC_2022, OFDM_VLC_2020}. Particularly, in the latter case,  the authors in \cite{PS_VLC_2022} proposed an adaptive spatial modulation (SM) scheme with PCS to improve the transmission rate of the VLC systems. The authors in \cite{OFDM_VLC_2020} proposed a probabilistically shaped orthogonal frequency multiplexed (OFDM) modulation for the VLC systems to approach the channel capacity and improve the symbol error performance. However, the existing studies focused only on utilizing the PCS scheme in point-to-point (P2P) communications with one-user scenarios. 

On the other hand, due to the broadcast nature of visible light signals, VLC systems can be categorized as broadcast networks. By using multiple light-emitting diode (LED) transmitters, VLC systems are capable of simultaneously serving multiple users (MU) by exploiting the spatial degrees of freedom (DoF) with precoding techniques. Several studies have examined the precoding designs for multi-user VLC systems with design objectives as sum-rate and weighted sum-rate \cite{ZF_2017},\cite{RSMA_VLC_2020}. It is important to note that, to facilitate the precoding design, these works considered a closed-form lower bound on the channel capacity, which was derived by assuming the continuous uniform distribution of the input. 
As a result, the shaping gain promised by PCS was not taken into consideration.  
To further improve the overall sum-rate performance, it is thus essential to exploit both the diversity and shaping gain by jointly optimizing the precoding matrix and the PCS. However, due to the presence of multi-user interference (MUI), a joint design of PCS and precoding for the multi-user VLC transmission presents significant challenges as the optimal design problem is shown to be complex and non-convex. 

Against the above background, this paper introduces a joint design of precoding and PCS for multi-user VLC systems to maximize the sum-rate performance where the data symbols are drawn from a probabilistically shaped $M$-ary pulse amplitude modulation ($M$-PAM). The design problem is shown to be non-convex, which renders solving the global optima challenging. To address the non-convex problem, we \textbf{first} present a firefly algorithm (FA) approach to solve a locally optimal solution to constellation probability distribution and precoding matrix simultaneously. \textbf{Second}, by adopting a particular precoding criterion, i.e., zero-forcing (ZF) precoding, we propose a low-complexity joint design, which is solved using an alternating optimization (AO) approach. 
\vspace{-0.25 cm}
\section{System Model}
\label{sec:model}

\subsection{VLC channel model}
\begin{figure}[ht]
    \centering
    \includegraphics[width = 8.8 cm]{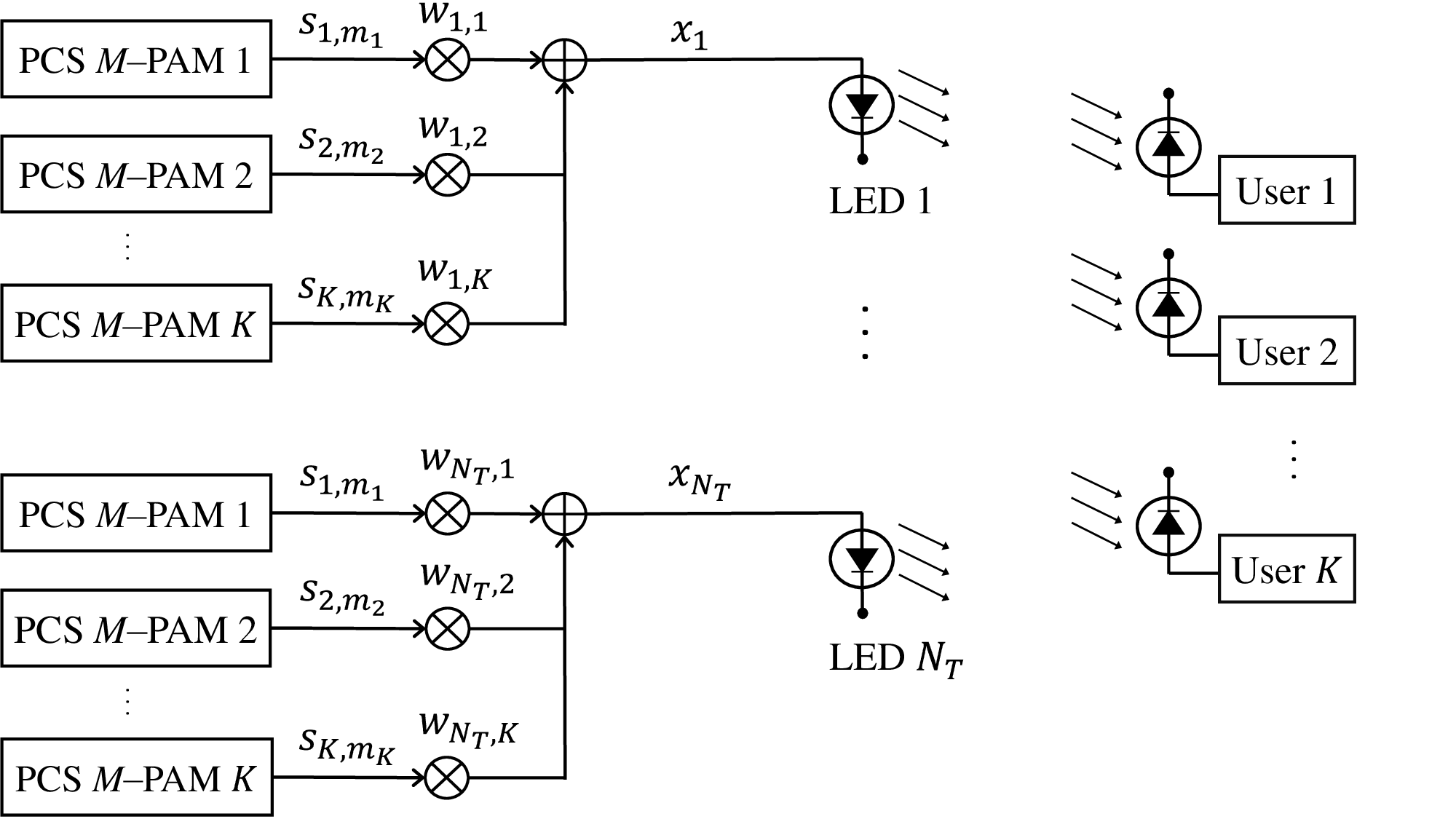}
    \caption{Schematic diagram of a multi-user VLC system with precoding and PCS.}
    \label{fig:system_model}
    \vspace{-0.25 cm}
\end{figure}

In indoor environments, the received VLC signal consists of two main components: the light-of-sight (LoS) and non-line-of-sight (NLoS) components due to reflections off ceilings and walls. It has been, however, shown in \cite{VLC_fundamental_2004} that the LoS signal accounts for the majority of the optical power received by the receiver, contributing to over 95\% of the total received optical power. For the sake of simplicity, we assume only the LoS propagation in this study.
The LoS channel coefficient $h$ between an LED luminary and the receiver can be given as
\begin{align}
\label{eqn:chann_coeff}
    h = \frac{\eta \gamma A_r}{d^2}L(\phi)T_s(\psi)g(\psi)\cos(\psi) \times \mathds{1}_{[0, \Psi]}(\psi), 
\end{align}
where $\eta$ is the LED electrical-to-optical conversion factor, $\gamma$ is the PD responsivity, $\mathds{1}_{[a,b]}(\cdot)$ represents the indicator function, $A_r$ is the active area of the photo-diode (PD) at the receiver, $d$ is the link length between the LED luminary and the PD, $\psi$ is the optical field of view (FOV) of the PD \cite{VLC_fundamental_2004}, and $\phi$ is the angle of incidence. In addition, $L(\phi) = \frac{l+1}{2\pi}\cos{(\phi)}$ is the emission intensity where $l=-\frac{\ln(2)}{\ln(\Theta_{0.5})}$ with $\Theta_{0.5}$ is the LED's semi-angle for half illuminance. $T_s(\psi)$ represents the optical filter gain and $g(\psi) = \frac{\kappa^2}{\sin^2(\Psi)} \times \mathds{1}_{[0, \Psi]}(\psi)$ is the optical concentrator gain. 


\subsection{Signal Model}
Our considered multi-user VLC system, as illustrated in Fig.~\ref{fig:system_model}, consists of $N_T$ LED arrays, $K$ independent users where each user is equipped with a single photo-diode (PD) receiver.

Let $\mathbf{s} = \begin{bmatrix}s_1 & s_2 & \cdots & s_K\end{bmatrix}^T \in \mathbb{R}^{K \times 1}$ be the vector of data symbols for $K$ users. Each data symbol $s_k$ is generated from a bipolar PCS $M$-PAM $k$-th modulation with $k = 1,\ 2, \ \cdots,\ K$. 
For the $k$-th PCS $M$-PAM  modulation, denote the vector of $M$ equally spaced bipolar $M$-PAM symbols as $\mathbf{\widehat{s}_k} = \begin{bmatrix} s_{k,1} \ s_{k,2} \ \cdots \ s_{k,M} \end{bmatrix} = \begin{bmatrix}s_{k,m_k}\end{bmatrix}_{\left(m_k = 1,\ 2,\ \cdots,\ M\right)}$ with 
the corresponding amplitude vector is  $\mathbf{a}_k = [ a_{k,1} \ a_{k,2} \ \cdots \ a_{k,M} ]= \begin{bmatrix}a_{k,m_k}\end{bmatrix}_{\left(m_k = 1,\ 2,\ \cdots,\ M\right)}$
and the corresponding probability mass function (PMF) vector is $\mathbf{p}_k = \begin{bmatrix} p_{k,1} \ p_{k,2} \ \cdots \ p_{k,M} \end{bmatrix}= \begin{bmatrix}p_{k,m_k}\end{bmatrix}_{\left(m_k = 1,\ 2,\ \cdots,\ M\right)}$. 
 
Denote $\mathbf{W} = \begin{bmatrix}\mathbf{w}_1 \ \mathbf{w}_2 \ \cdots \ \mathbf{w}_K \end{bmatrix}~ \in ~ \mathbb{R}^{N_T \times K}$ is the precoding matrix where $\mathbf{w_k} = \begin{bmatrix}w_{1,k} \ w_{2, k} \ \cdots \ w_{N_T, k}\end{bmatrix}^T$ is the $k-$th user's precoder vector with $w_{n,k}$ is the coefficient of transmitted signal from the $n$-th LED to $k$-th user for $n = 1,\ 2,\ \cdots, \ N_T$ and $k = 1,\ 2,\ \cdots, \ K$. Therefore, at the $n$-th LED transmitter, the broadcast signal $x_n$, which consists of data symbols of all users, can be given as
\begin{align}
\label{eqn: x_n}
    x_n = [\mathbf{W}]_{n,:} \times \mathbf{s},
\end{align}
where $[\mathbf{W}]_{n,:}$ is the $n$-th row of $\mathbf{W}$, which represents the precoder for $n$-th LED transmitter.
For illumination, a DC bias $I_{n}^{\text{DC}}$ is added to $x_n$ to generate a non-negative drive current $z_n$. In addition, to guarantee the operation of LEDs, i.e., to avoid the overheating problem and the potential light intensity reduction, $z_n$ must also be limited to a maximum threshold denoted by $I_{\text{max}}$. Therefore, we have $0 \leq z_n=x_n+I_n^{\text{DC}} \leq I_{\text{max}}$.
From \eqref{eqn: x_n} and denote $A = I_{\text{max}} - I_n^{\text{DC}}$, we have $0 \leq  [\mathbf{W}]_{n,:} \times \mathbf{s} \leq A$.

Let us assume that for all PCS $M$-PAM $k$ modulation with $k = 1,\ 2, \ \cdots,\ K$, the symbol peak amplitudes are the same, and denoted as $A$, i.e., $|a_{k,m_k}|~\leq~ A$. Due to the symbol symmetry around 0, for each PCS $M$-PAM modulation, the symbol amplitude levels can be given as $a_{k,m_k} = (2m_k-M-1)\frac{A}{M-1}$ for $m_k = 1,\ 2,\ \cdots, \ M$. 
Because the symbol amplitudes are in the range of $[-A,~A]$, to satisfy the peak amplitude constraint, the following constraint must be imposed
\begin{align}
    \left\lVert[\mathbf{W}]_{n,:}\right\rVert_1 \ \leq \ 1,\ \forall{n} \ \in \ \{1,\ 2,\ \cdots,\ N_T\}.
\end{align}

Let $\mathbf{H} = \begin{bmatrix} \mathbf{h}_1 & \mathbf{h}_2 & \cdots & \mathbf{h}_K\end{bmatrix}^{\text{T}}$  denote the channel matrix, where $\mathbf{h}_k = \begin{bmatrix} h_{1,k} & h_{2,k} & \cdots & h_{N_T,k}\end{bmatrix} ^\text{T} \in \mathbb{R}^{N_T \times 1}$ is the $k$-th user's channel vector. $h_{n,k}$ is the LoS channel coefficient between the $n$-th LED transmitter and the $k$-th user. At the user $k$-th, the received optical signals are captured by the PD and transformed into the electric signal as
\begin{align}
    \label{eqn:received_signal}
    y_k &= \mathbf{h}_k^T \mathbf{w}_k s_k + \underbrace{\mathbf{h}_k^T \sum_{i=1,\ i \neq k}^K \mathbf{w}_i s_i + n_k}_{\overline{y}_k},
\end{align}
where $n_k \ \sim \ \mathcal{N}(0, \ \sigma^2) $ is the additive white Gaussian noise (AWGN) with the power of $\sigma^2$. 

\vspace{-0.15 cm}
\section{Sum-rate maximization}
\begin{figure*}[h]
    \begin{align}
    \label{eqn:f_yk}
        f(y_k) = \sum_{p, \ (m_1, \ \cdots,\ m_K)_p \in A} \left[\prod_i^{K} \text{P}(s_i = s_{i,m_i}) \frac{1}{\sqrt{2\pi\sigma^2}} \exp{\left(-\frac{\left(y_k - \mathbf{h}^T_k \sum_{i=1}^K \mathbf{w}_i a_{i,m_i}\right)^2}{2\sigma^2}\right)}\right],
    \end{align}
    where $A = \{m_1, \ \cdots,\ m_K\} \times\{1, \ 2,\ \cdots,\ M\}$ is the Cartesian product of two sets and $(m_1, \ \cdots,\ m_K)_p$ is element $p$-th of set $A$. 
    \begin{align}
    \label{eqn:f_yk'}
        f(\overline{y}_k) = \sum_{q, \ (m_1, \ \cdots,\ m_K)_q \in A_k} \left[\prod_{j \neq k}^{K} \text{P}(s_j = s_{j, m_{j}}) \times \frac{1}{\sqrt{2\pi\sigma^2}} \exp{\left(-\frac{\left(\overline{y}_k - \mathbf{h}^T_k \sum_{j=1,j\neq k}^K \mathbf{w}_j a_{j,m_j}\right)^2}{2\sigma^2}\right)}\right],
    \end{align}
    where $A_k = \{\{m_1, \ \cdots,\ m_K\} \setminus{\{m_k\}\} \times\{1, \ 2,\ \cdots,\ M\}}$ and $(m_1, \ \cdots,\ m_K)_q$ is element $q$-th of set $A_k$. 
    \line(1,0){\linewidth}
    \vspace{-1 cm}
\end{figure*}
At the user $k$-th, the achievable rate can be given as
\begin{align}
    & R_{k} = \mathbb{I}(s_k;y_k) = h(y_k) - h(y_k|s_k) = h(y_k)-h(\overline{y}_k),\nonumber \\
          & = -\int_{-\infty}^{+\infty}f(y_k)\log_2{f(y_k)}\text{d}y_k + \int_{-\infty}^{+\infty}f(\overline{y}_k)\log_2{f(\overline{y}_k)}\text{d}\overline{y}_k,
\end{align}
where $\mathbb{I}(\cdot; \cdot)$ is the mutual information, $h(\cdot)$ and $h(\cdot|\cdot)$ are the differential and conditional entropy, respectively.
$f(y_k)$ is the probability density function of $y_k$, due to \eqref{eqn:received_signal}, $f(y_k)$ is given by \eqref{eqn:f_yk} and $f(\overline{y}_k)$ is given by \eqref{eqn:f_yk'}.

With pre-defined symbol amplitude levels in each PCS-PAM constellation i.e., $a_{k,m_k}$, the achievable rate $R_k$ is parameterized with these variables: precoding matrix $\mathbf{W}$ and probability mass function (PMF) symbol vectors of $K$ PCS $M$-PAM modulations, i.e., $\mathbf{p}_1,\ \mathbf{p}_2, \ \cdots, \ \mathbf{p}_K$. Denote $\mathbf{P} = \begin{bmatrix}\mathbf{p}_1 & \mathbf{p}_2 & \cdots & \mathbf{p}_K\end{bmatrix}^T \in \mathbb{R}^{K \times M}$ is used for mathematical convenience in the later parts of the paper. 
Therefore, the sum-rate maximization problem can be formulated as
\begin{subequations}
\vspace{-0.3 cm}
    \label{original_problem}    
    \begin{alignat}{2}
        \mathbb{P} \mathbf{1}:~
        &\underset{\mathbf{P}, \mathbf{W}}{\text{maximize}} \ \ \sum_{k=1}^K R_k (\mathbf{P}, \mathbf{W})
        \label{eqn: objectiveP1}\\ 
        &\text{subject to } \ \ \ \left\lVert[\mathbf{W}]_{n,:}\right\rVert_1 \ \leq \ 1,\label{eqn:constraint_peak}\\
        & \hspace{1.8 cm} \mathbf{0}_{K \times M} \ < \mathbf{P}  < \ \mathds{1}_{K \times M} , \label{eqn:constraint_P1}\\
        & \hspace{1.8 cm}  \mathbf{P} \times \mathds{1}_{M\times 1} = \mathds{1}_{K\times 1},\label{eqn:constraint_P2}
    \end{alignat}    
\end{subequations}
where \eqref{eqn: objectiveP1} is the sum rate formula and \eqref{eqn:constraint_peak} is the peak power constraint for LEDs. It can be observed that problem $\mathbb{P} \mathbf{1}$ is non-convex due to the non-convexity of the objective function. Thus, it is generally difficult to optimally solve it. 
Motivated by the above observations, we introduce a novel Firefly algorithm (FA) approach to simultaneously solve $\mathbf{P}$ and $\mathbf{W}$ for the original $\mathbb{P} \mathbf{1}$ in the following sections.


\subsection{Proposed FA Approach}
Adopting the penalty method as \cite{natural_meta_2014}, problem $\mathbb{P} \mathbf{1}$ can be equivalently reformulated as
\begin{align}
\vspace{-0.5 cm}
    \mathbb{P} \mathbf{2}:~
    &\underset{\mathbf{P}, \mathbf{W}}{\text{maximize}} \ \ \sum_{k=1}^K R_k(\mathbf{P}, \mathbf{W})  - P(\mathbf{P}, \mathbf{W}),
    \label{objectiveP2}
\end{align}
where $P(\mathbf{P}, \mathbf{W})$ is the penalty term given as 
\begin{align}
    \label{eqn:penalty_term}
    &P(\mathbf{P}, \mathbf{W}) = 
    \lambda_1 \sum_{n=1}^{N_t}\max{\left(0, ~\left\lVert[\mathbf{W}]_{n,:}\right\rVert_1 - 1\right)}^2\\\nonumber
    & + \lambda_2 \sum_{k=1}^{K}\sum_{m=1}^M \min{\left(0,~ p_{k,m}\right)}^2 + \lambda_3 \sum_{k=1}^{K}\sum_{m=1}^M \max{\left(0,~ p_{k,m} - 1\right)}^2 \\\nonumber
    & + \lambda_4 \sum_{k=1}^K \max{\left(0, ~\left\lVert[\mathbf{P}]_{k,:}\right\rVert_1 - 1\right)}^2, \ \ \lambda_j \ \text{are penalty constants.}
\end{align}

The Firefly algorithm (FA) was proposed based on the firefly behaviors with three idealized rules \cite{natural_meta_2014}, \cite{FA_2009}. First, all fireflies are unisex, so one firefly will be attracted to other fireflies regardless of their sex. Second, the attractiveness of any firefly to the other one is proportional to its brightness, and both decrease as their distance increases. For any two flashing fireflies, the less bright one will move towards the brighter one. If there is no brighter one than a particular firefly, it will move randomly. Third, the brightness of a firefly is determined by the landscape of the objective function. 

Let $(\mathbf{W}_n, \mathbf{P}_n)$ be the particular location of firefly $n$-th amongst the population of $N$ fireflies, i.e., $n \in \{1,2, \cdots, N\}$. Since the proposed optimization problem is a maximization, the brightness of the  firefly $n$-th is determined as the value of the objective function at $(\mathbf{W}_n, \mathbf{P}_n)$ as
\begin{align}
    \label{eqn:light_intensity}
    I(\mathbf{W}_n, \mathbf{P}_n) = \sum_{k=1}^K R_k(\mathbf{W}_n, \mathbf{P}_n)  - P(\mathbf{W}_n, \mathbf{P}_n).
\end{align}

For any fireflies $m$-th and $n$-th amongst the population in the generation $t$. If $I\left(\mathbf{W}^{(t)}_n, \mathbf{P}^{(t)}_n\right) > I\left(\mathbf{W}^{(t)}_m, \mathbf{P}^{(t)}_m\right)$, the firefly $m$-th will move toward the firefly $n$-th as
\begin{align}
\nonumber
    &\mathbf{W}_m^{(t+1)} \!=\! \mathbf{W}_m^{(t)} \!+\! \beta_0 \exp{\left(\!-\!\gamma\left(r^{(t)}_{\mathbf{W},mn}\right)^2\right)} \left(\mathbf{W}^{(t)}_n \!-\! \mathbf{W}^{(t)}_m\right) \!+\! \alpha^{(t)} \mathbf{V_W},\\
    &\mathbf{P}_m^{(t+1)} \!=\! \mathbf{P}_m^{(t)} \!+\! \beta_0 \exp{\left(\!-\!\gamma\left(r^{(t)}_{\mathbf{P},mn}\right)^2\right)} \left(\!\mathbf{P}^{(t)}_n \!-\! \mathbf{P}^{(t)}_m\!\right) \!+\! \alpha^{(t)} \mathbf{V_P}, \label{eqn:update_PW}
\end{align}
where $r^{(t)}_{\mathbf{W},mn} = \left\lVert \mathbf{W}^{(t)}_n - \mathbf{W}^{(t)}_m\right\rVert$ and $r^{(t)}_{\mathbf{P},mn} = \left\lVert \mathbf{P}^{(t)}_n - \mathbf{P}^{(t)}_m\right\rVert$ are the Cartesian distances, $\beta_0$ is the attractiveness at $r^{(t)}_{\mathbf{W},mn} = 0$ and $r^{(t)}_{\mathbf{P},mn} = 0$. $\gamma$ is the variation of attractiveness, $\alpha^{(t)} = \alpha_0^t$ is the random factor at generation $t$ with $\alpha_0$ is the initial random factor and $V_W ~ \in ~\mathbb{R}^{N_T \times K}$, $V_P ~ \in ~\mathbb{R}^{K \times M}$ are random matrices whose elements are drawn from a normal distribution. The proposed Firefly algorithm is summarized in Algorithm \ref{alg1}. 
\begin{algorithm}[ht]
\small
    \caption{ Firefly algorithm for solving $\mathbb{P} \mathbf{2}$}\label{alg1}
    \begin{algorithmic}[1]
    \STATE \textbf{Input}: Population size $N$, maximum generation $T$.
    \STATE Generate $N$ populations $\{\left(\mathbf{W}_1, \mathbf{P}_1\right),  \cdots, \left(\mathbf{W}_N, \textbf{P}_N\right)\}$ randomly.
    \STATE Evaluate the light intensities of $N$ population $I(\mathbf{W}_i, \mathbf{P}_i) ~ \forall{i ~ \in ~ [1, \ N]}$ as \eqref{eqn:light_intensity}.
    \STATE Rank the fireflies in descending order of light intensities $I(\mathbf{W}_i, \mathbf{P}_i)$.
    \STATE Define the current best solution: $(\mathbf{W}^*,\mathbf{P}^*) \leftarrow (\mathbf{W}_1, \mathbf{P}_1), \ I^* \leftarrow I(\mathbf{W}^*, \mathbf{P}^*)$.
            \FOR{$t = 1~:~T$}
                \FOR{$m = 1~:~N$} 
                    \FOR{$n = 1~:~N$} 
                        \IF{$I(\mathbf{W}_n, \mathbf{P}_n) > I(\mathbf{W}_m, \mathbf{P}_m)$} 
                             \STATE 1. Move firefly $m$ toward firefly $n$ as \eqref{eqn:update_PW}.
                             \STATE 2. Update the light intensity of firefly $m$ with new $\left(\mathbf{W}_m, \mathbf{P}_m\right)$ as \eqref{eqn:light_intensity}.
                        \ENDIF
                    \ENDFOR
                \ENDFOR 
                \STATE Rank the fireflies in descending order of $I(\mathbf{W}_i, \mathbf{P}_i)$.
                \STATE Update the current best solution $(\mathbf{W}^*, \mathbf{P}^*) \leftarrow (\mathbf{W}_1, \mathbf{P}_1), \ I^* \leftarrow I(\mathbf{W}^*, \mathbf{P}^*)$.
            \ENDFOR
    \STATE Return the solution $(\mathbf{W}^*, \mathbf{P}^*)$.
    \label{alg.1}
    \end{algorithmic}
\end{algorithm}
It is important to note that according to \cite{FA_2009}, FA can effectively find the global optima and all the local optima. It tends to be a global optimizer but at the potential expense of large computational complexity.

\subsection{Low-complexity Design with Zero-forcing Precoding}
To avoid the computational complexity of the joint design probabilistic shaping and precoding based on the FA approach, we employ the suboptimal zero-forcing (ZF) precoding strategy. The assumption is applicable as the CSI is practically available at the transmitters in the local network setting of VLC systems. Furthermore, while the ZF offers a lower complexity, its sum-rate performance can be further improved in the MU-VLC systems when additional transmitters can be deployed.

With ZF precoding, the MUI at each user's received signal $y_k$ is completely removed via the construction of precoder $\mathbf{w_k}$ in such a way that is orthogonal to the channel vectors of other users, i.e., $\mathbf{h}_k \mathbf{w}_i = 0, \ \forall{i \neq k}$ \cite{ZF_2017}. The orthogonality of $\mathbf{W}$ results in lower computational complexity at the expense of decreased performance since the degrees of freedom are reduced compared to the general case.  
By removing the MUI via ZF precoding, the received electrical signal at user $k$-th is simplified to
$y_k =  \mathbf{h}_k^T \mathbf{w}_k s_k + n_k$.
As a result, the achievable rate at user $k$-th can be given as
\begin{align}
    R_{k} & = \mathbb{I}(s_k;y_k)  
            = h(y_k) - h(y_k|s_k) = h(y_k) - h(n_k) \\\nonumber
          & = -\int_{-\infty}^{+\infty}f(y_k)\log_2{f(y_k)}\text{d}y_k -\frac{1}{2}\log_2(2\pi e \sigma^2),
\end{align}
where $f(y_k)$ is the probability density function of $y_k$, given by
\begin{align}
    f(y_k) =\sum_{m_k}^{M} p_{k,m_k} \frac{1}{\sqrt{2\pi\sigma^2}} \exp{\Big(-\frac{\left(y_k - \mathbf{h}^T_k \mathbf{w}_k a_{k,m_k}\right)^2}{2\sigma^2}\Big)},  
\end{align}
with $k = 1,\ 2, \ \cdots,\ K$, $m_k = 1,\ 2,\ \cdots,\ M$, $p_{k,m} = \text{P}(s_k = s_{k,m_k})$.

By quantizing the continuous source $y_k$ with a sufficiently small step size $\Delta$, the achievable rate $R_k$ can be alternatively derived by a Riemann sum with $N$ rectangular partitions whose width is $\Delta$ and height is $f(y_k^n)\log_2{f(y_k^n)}$, as follows:
\begin{align}
    R_{k} &= -\sum_{n=1}^N f(y_k^n)\log_2{f(y_k^n)}\Delta -\frac{1}{2}\log_2(2\pi e \sigma^2)\\\nonumber
    &= -\sum_{n=1}^N \left(\frac{1}{\sqrt{2\pi \sigma^2}} \sum_{m_k}^M p_{k,m_k} \exp{\left(-\frac{\left(y_k^n - \mathbf{h}_k^T \mathbf{w}_k a_{k,m_k}\right)^2}{2\sigma^2}\right)}\right)\\\nonumber
    &\times \log_2{\left(\frac{1}{\sqrt{2\pi \sigma^2}} \sum_{m_k}^M p_{k,m_k} \exp{\left(-\frac{\left(y_k^n - \mathbf{h}_k^T \mathbf{w}_k a_{k,m_k}\right)^2}{2\sigma^2}\right)}\right)}\Delta \\\nonumber
    & - \frac{1}{2}\log_2(2\pi e \sigma^2).
\end{align}
Then, the sum-rate maximization problem in the case of using ZF precoding can be formulated as
\begin{subequations}
\vspace{-0.3 cm}
    \label{ZF_problem}    
    \begin{alignat}{2}
        \mathbb{P} \mathbf{3}:~
        &\underset{\mathbf{P}, \mathbf{W}}{\text{maximize}} \ \ \sum_{k=1}^K R_k (\mathbf{P}, \mathbf{W})
        \label{eqn: objectiveP3}\\ 
        &\text{subject to } \ \ \      
        \mathbf{h}_k \mathbf{w}_i = 0, \ \ \forall{i \neq k},\label{eqn:constraint_ZF}\\
        & \hspace{1.9 cm} \left\lVert[\mathbf{W}]_{n,:}\right\rVert_1 \ \leq \ 1,\label{eqn:constraint_peak_ZF}\\
        & \hspace{1.9 cm} \mathbf{0}_{K \times M} \ < \mathbf{P}  < \ \mathds{1}_{K \times M} , \label{eqn:constraint_P1_ZF}\\
        & \hspace{1.9 cm}  \mathbf{P} \times \mathds{1}_{M\times 1} = \mathds{1}_{K\times 1},\label{eqn:constraint_P2_ZF}
    \end{alignat}    
\end{subequations}
Due to the non-convexity of the objective function, $\mathbb{P} \mathbf{3}$ is a non-convex 
problem with two optimization variables $\mathbf{P}$ and $\mathbf{W}$. Thus, it is generally difficult to solve it. To cope with this situation, as $\mathbf{P}$ and $\mathbf{W}$ are two independent variables, they can be alternatively solved with an alternating optimization (AO) approach \cite{AO_2008}, and a sub-optimal solution can be obtained. 
The procedure of the AO approach is presented as follows:

1) Beginning with an initial value of precoding matrix $\mathbf{W}^{(0)}$, the following sub-problem with respect to PMF matrix $\mathbf{P}$ will be solved at the $r$-th iteration:
    \begin{subequations}
    \vspace{-0.25 cm}
    \label{ZF_problem_P}    
        \begin{alignat}{2}
            \mathbb{P} \mathbf{3(a)}:~
            &\underset{\mathbf{P}}{\text{maximize}} \ \ \sum_{k=1}^K R_k (\mathbf{P}, \mathbf{W}^{(r-1)})
            \label{eqn: objectiveP3a}\\ 
            &\text{subject to } \ \ \       
             \mathbf{0}_{K \times M} \ < \mathbf{P}  < \ \mathds{1}_{K \times M}, \label{eqn:constraint_P1_ZF_P}\\
            & \hspace{1.9 cm}  \mathbf{P} \times \mathds{1}_{M\times 1} = \mathds{1}_{K\times 1}.\label{eqn:constraint_P2_ZF_P}
        \end{alignat}    
    \end{subequations}
    It is seen that $\mathbb{P} \mathbf{3(a)}$ is convex since the objective function is a concave function of $\mathbf{P}$ according to \cite{Elements_IT_2012} and the two constraints are linear. Therefore, it can be solved efficiently using standard optimization packages, such as CVX \cite{cvx}. 
    
2) The precoding matrix $\mathbf{W}^{(r)}$ is then updated from the optimal solution of  $\mathbb{P} \mathbf{3(a)}$ at $r$-th iteration, i.e., $\mathbf{P^{(r)}}$, by solving the following sub-problem:
    \begin{subequations}
    \vspace{-0.3 cm}
    \label{ZF_problem_W}    
        \begin{alignat}{2}
            \mathbb{P} \mathbf{3(b)}:~
            &\underset{\mathbf{W}}{\text{maximize}} \ \ \sum_{k=1}^K R_k (\mathbf{P}^{(r)}, \mathbf{W})
            \label{eqn: objectiveP3b}\\ 
            &\text{subject to } \ \ \       
           \mathbf{h}_k \mathbf{w}_i = 0, \ \ \forall{i \neq k},\label{eqn:constraint_ZF_W}\\
            & \hspace{1.7 cm} \left\lVert[\mathbf{W}]_{n,:}\right\rVert_1 \ \leq \ 1.\label{eqn:constraint_peak_ZF_W}
        \end{alignat}    
    \end{subequations}
    It is seen that $\mathbb{P} \mathbf{3(b)}$ is not convex due to the non-convexity of the objective function for $\mathbf{W}$.
    To tackle it, the convex-concave procedure (CCP) is employed to solve local optima \cite{yuille_CCP_2003}. Introducing the following slack variables $x_k^n$, $\forall{k = 1,\ \cdots, \ K}$ and $\forall{n = 1, \ \cdots, \ N}$, $\mathbb{P} \mathbf{3(b)}$ can be reformulated as
    \begin{subequations}
    \label{ZF_problem_W2}    
        \begin{alignat}{2}
            \label{eqn: objectiveP3b2}
            &\overline{\mathbb{P} \mathbf{3(b)}}:~
            \underset{\mathbf{W}}{\text{maximize}} \\\nonumber
            & \sum_{k=1}^K \sum_{n=1}^N \left(\frac{-1}{\sqrt{2\pi \sigma^2}} \sum_{m=1}^M p_{k,m}^{(r)} x_k^n\right) \log_2{\left(\frac{1}{\sqrt{2\pi \sigma^2}} \sum_{m=1}^M p_{k,m}^{(r)} x_k^n\right)}\Delta  \\\nonumber
            & - \frac{1}{2}\log_2(2\pi e \sigma^2) \\
            &\text{subject to } \ \ \       
            \label{eqn:constraint_slack_var2}x_k^n = \exp{\left(-\frac{\left(y_k^n - \mathbf{h}_k^T \mathbf{w}_k a_{k,m_k}\right)^2}{2\sigma^2}\right)}, \\\nonumber
            & \hspace{1.7 cm} \eqref{eqn:constraint_ZF_W}, \eqref{eqn:constraint_peak_ZF_W}.
        \end{alignat}    
    \end{subequations}
    It is seen that the objective function is concave and constraints \eqref{eqn:constraint_ZF_W} and \eqref{eqn:constraint_peak_ZF_W} are convex, but \eqref{eqn:constraint_slack_var2} is not convex. The first-order Taylor approximations are applied to approximate this non-convex constraint to its respective linear bounds to cope with this situation. Accordingly, $\overline{\mathbb{P} \mathbf{3(b)}}$ can be reformulated as
    \begin{subequations}
    \vspace{-0.5 cm}
    \label{ZF_problem_W3}    
        \begin{alignat}{2}
            \label{eqn: objectiveP3b3}
            &\widehat{\mathbb{P} \mathbf{3(b)}}:~ \nonumber
            \underset{\mathbf{W}}{\text{maximize}} \hspace{0.5 cm}\eqref{eqn: objectiveP3b2} \\
            & \text{subject to } \ \ \    
             x_k^n = 
            \exp{\left(-\frac{z_{\{n, j-1\}}^2}{2\sigma^2}\right)} \\\nonumber 
            & \hspace{1 cm} + \exp{\left(-\frac{z_{\{n, j-1\}}^2}{2\sigma^2}\right)} \frac{z_{\{n, j-1\}}}{\sigma^2} \mathbf{h}_k^T \left(\mathbf{w}_k - \mathbf{w}_k^{(j-1)}\right)a_{k,m_k},\label{eqn:constraint_slack_var3} \\\nonumber
            & \hspace{1 cm} \eqref{eqn:constraint_ZF_W}, \eqref{eqn:constraint_peak_ZF_W}. 
        \end{alignat}
    \end{subequations}
    where $z_{\{n, j-1\}} = y_k^n - \mathbf{h}_k^T \mathbf{w}_k^{(j-1)} a_{k,m_k}$, and $\mathbf{w}_k^{(j-1)}$ are the solutions of $\mathbf{w}_k$ at the $(j-1)$-th iteration of the convex-concave procedure. Problem $\widehat{\mathbb{P} \mathbf{3(b)}}$ is now convex and can be efficiently solved with the CCP algorithm Algorithm 2. 
\begin{algorithm}[ht]
\small
\label{alg2}
    \caption{\small CCP algorithm for solving $\overline{\mathbb{P} \mathbf{3(b)}}$}
    \begin{algorithmic}[1]
    \STATE \textbf{Input}: Maximum number of iterations ${N}_{\text{max}}$, the error tolerance $\epsilon$ 
    \STATE Generate the feasible starting point $\mathbf{W}^{(0)}$.
    \WHILE{convergence = \textbf{False} and $j \leq {N}_{\text{max}}$} 
    \STATE Solve $\widehat{\mathbb{P} \mathbf{3(b)}}$ using $\mathbf{W}^{(j-1)}$ from the $(j-1)$-th iteration. 
    \STATE \IF{$\frac{\norm{\mathbf{W}^{(j)} - \mathbf{W}^{(j-1)}}}{\norm{\mathbf{W}^{(j)}}}\leq \epsilon$} 
    \STATE convergence = \textbf{True}
    \STATE $\mathbf{W}^* \leftarrow \mathbf{W}^{(j)}$ \ELSE
    \STATE convergence = \textbf{False} \ENDIF
    \STATE $j \leftarrow j+1 $
    \ENDWHILE   
    \STATE Return the solution $\mathbf{W}^*$
    \end{algorithmic}
\end{algorithm}

The AO approach repetitively solves $\mathbb{P} \mathbf{3(a)}$ and $\mathbb{P} \mathbf{3(b)}$ in $N_0$ iterations to obtain the solution for $\mathbb{P} \mathbf{3}$.

\vspace{-0.1 cm}
\section{Results and Discussions}

This section presents simulation results to illustrate the effectiveness of the proposed joint design probabilistic shaping and precoding in our considered multi-user VLC system. 
For indoor configuration, a room measuring $(5 \text{m} \times 5 \text{m} \times 3 \text{m})$ as (Length $\times$ Width $\times$ Height) with four LED luminaries respectively positioned at $(\pm \sqrt{2}, \pm \sqrt{2}, 3)$ is considered. The positions of user 1 and user 2 are $(1.25, -1.6, 0.5)$ and $(-2.25, -0.33, 0.5)$ respectively. 
For LEDs and receiver photodetectors, $\eta = 0.44 \ \text{W/A}, \gamma = 0.54 \ \text{A/W}, \Theta_{0.5} = 60 ^{\circ}$, $A_r = 1 \text{cm}^2$, $\Psi = 60 ^{\circ}$, $T_s(\psi) = 1$ and $\kappa = 1.5$. For FA, $\gamma = 1, \ \beta_0 = 1,\ \alpha_0 = 0.9$, $N = 100$, $T = 35$ and $\lambda_j = 10^4$.   
\vspace{-0.3 cm}
\begin{figure}[h!]
    \centering
    \includegraphics[width = 8.8 cm]{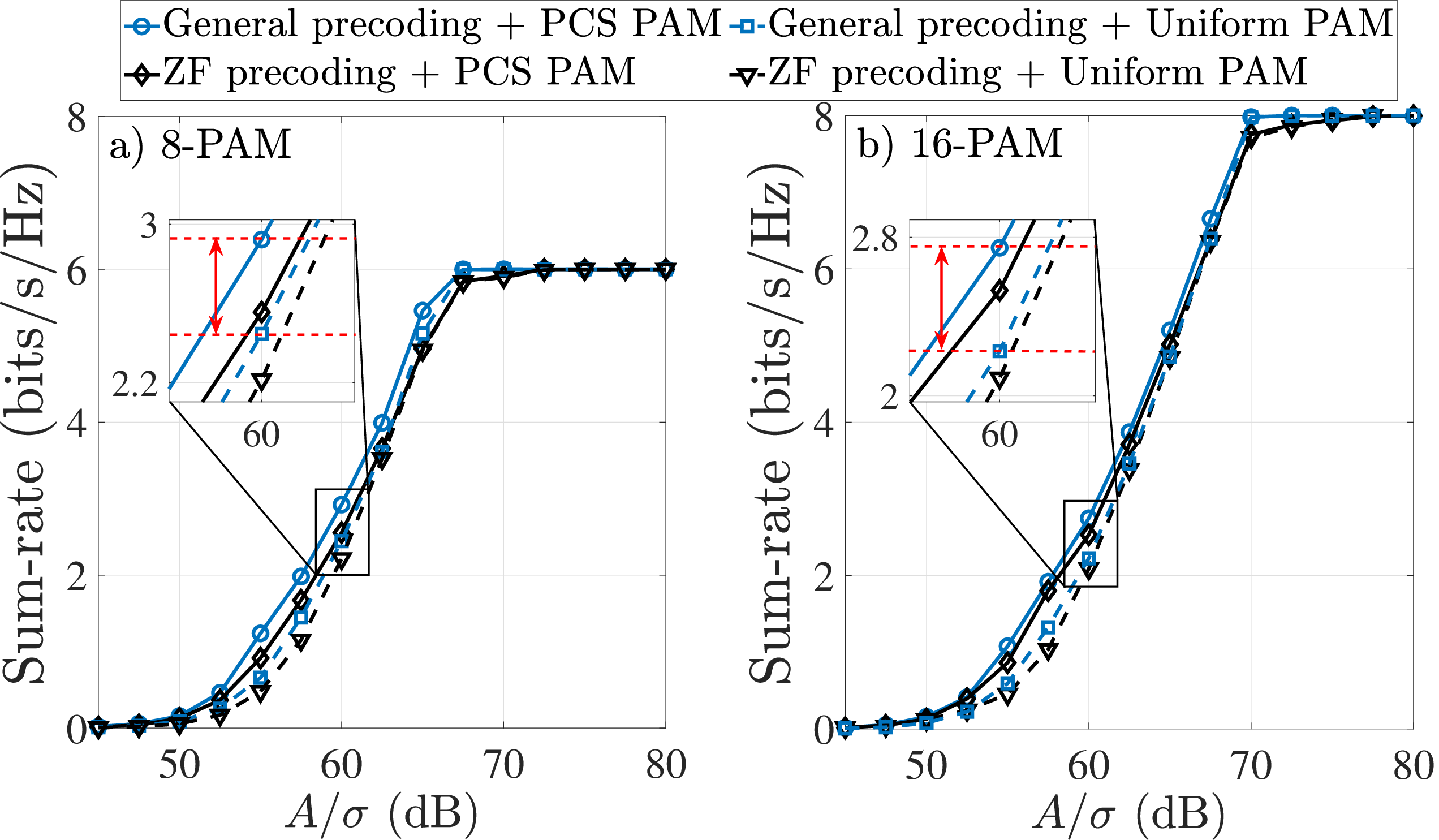}
    \caption{Maximal sum rate versus $A / \sigma$.}
    \label{fig:SR_A}
    \vspace{-0.3 cm}
\end{figure}
\begin{figure*}[ht]
    \centering
    \includegraphics[width = 17.5 cm]{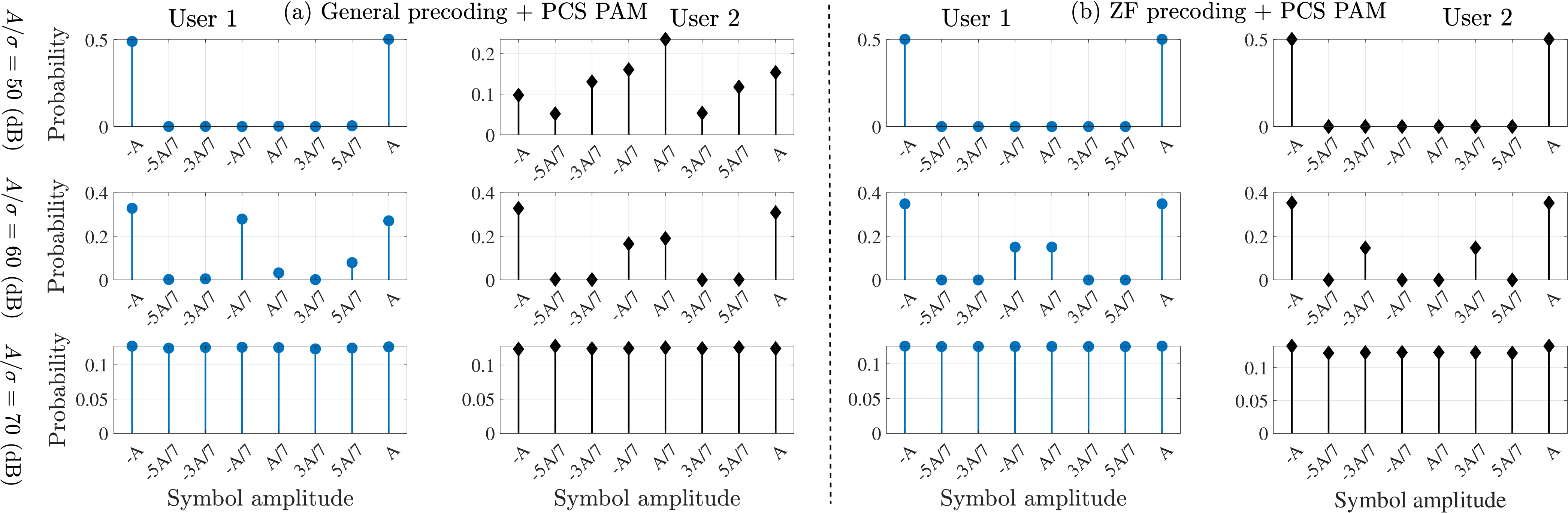}
    \caption{Optimal symbol distributions of PCS 8-PAM with different $A / \sigma$.}
    \label{fig:symbol} 
    \vspace{-0.7 cm}
\end{figure*}

Fig. \ref{fig:SR_A} shows the sum-rate performance system obtained by the proposed joint design and conventional design with uniformly distributed PAM versus $A / \sigma$. We consider two approaches, one using the FA to find the optimal design with general precoding and the other using the AO to find the sub-optimal solution with ZF precoding. As seen, the proposed design can provide a better sum-rate value than the scheme with uniform signaling. For example, at $A / \sigma = 60$ dB, the proposed joint design significantly outperforms the uniformly distributed scheme by approximately 0.48 bits/s/Hz ($19.7\%$) for 8-PAM and 0.52 bits/s/Hz ($23.4\%$) for 16-PAM. At the higher $A / \sigma$, however, the rate gap between the proposed design and uniform distributed scheme becomes narrower. This is because, with optimal precoding matrix, MUI is almost eliminated, leaving only the entropy of the transmitted signal $\mathbf{h}_k^T\mathbf{w}_k s_k$ at the rate of user $k$-th, which is maximized by uniform input distribution at high $A / \sigma$. In addition, the rate gap between the design with general precoding and ZF precoding is realized because of the lower degrees of freedom in designing $\mathbf{W}$ with orthogonality constraint.

Fig. \ref{fig:symbol} shows the optimal symbol distribution of PCS 8-PAM with different $A / \sigma$. As seen, under a lower $A / \sigma$, only a few symbols are assigned with probabilities to reinforce the noise immunity. However, as  $A / \sigma$ increases, more symbols are used, and the symbol distribution becomes nearly uniform. From Fig. \ref{fig:SR_A}, at higher $A / \sigma$, this transition to uniform distribution is evident where the rate gap between the proposed joint design and uniform distributed scheme is diminishing. Also, it is worth noting that, for the design with ZF precoding, the symbol distribution is symmetric around 0, and the number of active symbols increases when $A / \sigma$ increases. This behavior can be explained as follows. Through the implementation of ZF precoding, the MUI is fully removed, resulting in solely the entropy of the transmitted signal $\mathbf{h}_k^T\mathbf{w}_k s_k$ at the data rate of each user $k$-th. Thus, using $M$-PAM modulation with the symbol interval $[-A, A]$, this rate is maximized by a symmetric symbol distribution with a unique number of active points determined by the $A / \sigma$ value \cite{GM_2016}. Furthermore, different optimal symbol distributions can be observed for different users. 
\vspace{-0.3 cm}
\begin{figure}[ht]
    \centering
    \includegraphics[height = 4.5cm]{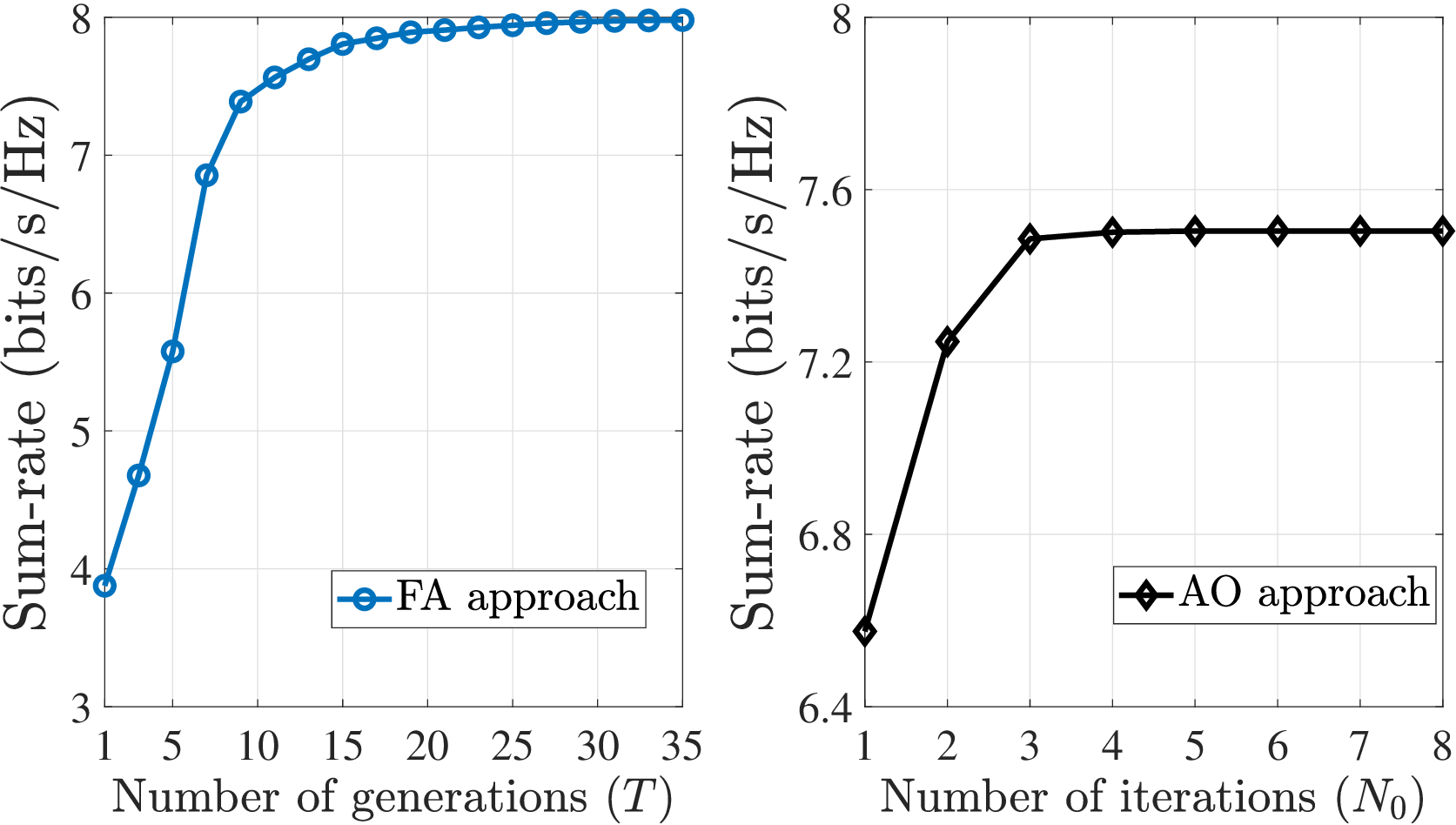}
    \caption{Convergence behaviors of FA and AO approach with PCS 16-PAM and $A / \sigma = 70$ dB.}
    \label{fig:convergence}
    \vspace{-0.4 cm}
\end{figure}

Finally, the convergence behaviors of the FA and AO algorithms are presented in Fig. \ref{fig:convergence} with PCS 16-PAM and $A / \sigma = 70$ dB. It is shown that the proposed FA-based design achieves a better performance than the AO-based approach at the expense of increased complexity. Particularly, the former attains its optimal solution after 30 generations, whereas the solution of the latter can be solved after only 5 iterations. 

\vspace{-0.25 cm}
\section{Conclusion}
This paper proposes a joint design of PCS and precoding to enhance the sum-rate performance of multi-user VLC broadcast channels subject to peak amplitude constraint. Two design approaches based on FA and AO are presented. Numerical simulations verified the superiority of the proposed joint design relative to the conventional design with uniform signaling. Furthermore, the optimal symbol distribution in terms of optical amplitude is also discussed. 
\bibliographystyle{IEEEtran}
\vspace{-0.25 cm}
\bibliography{references}
\end{document}